\documentclass[epsfig,aps,prl,twocolumn]{revtex4}
\usepackage{amssymb}
\usepackage{amsmath}
\usepackage{graphicx}
\usepackage{dcolumn}
\usepackage{bm}
\usepackage{epsfig}

\begin{document}
\newcommand{\eexp}[1]{\mbox{e}^{#1}}
\title{Decay of Quantum Correlations in Atom Optics Billiards with Chaotic and
Mixed Dynamics }
\author{M. F. Andersen, A. Kaplan, T. Gr\"{u}nzweig, and N. Davidson.}
\address{Department of Physics of Complex Systems, Weizmann Institute of Science,\\
Rehovot 76100, Israel}
\date{\today}

\begin{abstract}
We perform echo spectroscopy on ultra cold atoms in atom optics billiards, to study their quantum dynamics. The
detuning of the trapping laser is used to change the ``perturbation'', which causes a decay in the echo coherence.
Two different regimes are observed: First, a perturbative regime in which the decay of echo coherence is
non-monotonic and partial revivals of coherence are observed. These revivals are more pronounced in traps with
mixed dynamics as compared to traps where the dynamics is fully chaotic. Next, for stronger perturbations, the
decay becomes monotonic and independent of the strength of the perturbation. In this regime no clear distinction
can be made between chaotic traps and traps with mixed dynamics.
\end{abstract}
\pacs{}
\maketitle

The decoherence of superposition states, as they interact with the environment \cite{Zurek03}, is one of the main
obstacles in realizing quantum information processing schemes \cite{DiVincenzo95}. Decoherence or dephasing
represent loss of information, quantified by a decrease in the ``fidelity'', which typically measures the overlap
between a desired state and the actual output of a real system.

In the field of classical chaos exponential sensitivity to initial conditions (non-vanishing Lyapunov exponent)
plays a central role. However, the quantum analogue of evolving two initially slightly different wavefunctions in
the same Hamiltonian yields that their overlap is constant in time, thereby not giving any information on the
dynamics of the system. The fingerprints of the classically chaotic dynamics in the quantum regime (Quantum chaos)
is a rich field of study \cite{Gutzwiller90}. One approach is to study the decay of the overlap between two
initially identical wavefunctions evolved in slightly different Hamiltonians. The fidelity then denotes the
overlap between a state evolved by a Hamiltonian $H_{\uparrow }$ with the same state evolved by a slightly
perturbed Hamiltonian $H_{\downarrow }$ \cite{Peres84}. The fidelity is in this context often denoted the
Loschmidt Echo, since it is equivalent to the overlap between an initial state, and the same state evolved forward
in time in $H_{\uparrow }$ and then backwards in time in $H_{\downarrow }$ \cite{Jalabert01}. The decay of
fidelity in chaotic systems and its dependence on several parameters, has been the topic of intense theoretical
studies in recent years (see Ref. \cite{Peres84, Jalabert01, Emerson02} and references therein). Nevertheless,
experimental studies of chaotic systems are still lacking (mostly due to the difficulty of preparing
highly-excited pure quantum states) and so are both theoretical and experimental investigations of systems with
mixed dynamics. Of special interest in quantum dynamical studies are the ``cross-over'' regimes through which the
system goes by changing its parameters from those values for which a quantum description is required, to those
that allow classical models to be used. Understanding these regimes can shed light on the relation between
classical and quantum properties of the system.

Ultra cold atoms have been used in the past to experimentally study both quantum and classical dynamics. Quantum
dynamics have been studied in driven 1D systems, where a broad variety of phenomena such as dynamical localization
\cite{Raizen99}, dynamical tunneling \cite{Tunneling} and quantum accelerator modes \cite{Accelerator} have been
demonstrated. Classical dynamics has been studied in atom-optics billiards \cite{BilliardsRaizen, Billiards}, in
which regular, chaotic and mixed dynamics were observed.

In this letter we use microwave (MW)``echo spectroscopy'' \cite{Andersen03} to experimentally measure the
dephasing and quantum dynamics of ultra cold $^{85}$Rb atoms trapped in atom-optics billiards, with underlying
chaotic or mixed classical dynamics. Echo spectroscopy measures the overlap between two initially identical states
evolved in slightly different Hamiltonians (and is therefore closely related to the fidelity \cite{Andersen04b}),
but it overcomes the need for pure quantum states and allows the use of thermal ensembles for the study of quantum
dynamics \cite{Andersen03}. We demonstrate that the decay of the echo coherence displays qualitative different
behavior for different perturbation strengths. Two distinct regimes are observed. First, a ``perturbative'' regime
where the decay depends on the perturbation strength and where partial coherence (wavepacket) revivals are
observed even when the underlying classical dynamics is chaotic. This indicates that the partial revivals observed
for a nearly harmonic trap \cite{Andersen03} are in fact a generic feature of trapped atoms, as predicted in
\cite{Andersen04b}. For stronger perturbations we observe a crossover to a regime where the decay of the echo
coherence is independent of the perturbation strength. In this regime the decay is monotonic and no revivals are
observed. For traps where the classical motion exhibits a mixed phase space (i.e. stable "islands" in a chaotic
"sea") we observe more pronounced revivals in the perturbative regime, due to the periodic classical motion in the
islands. However, a perturbation-independent regime exists also for these systems, and then the decay is
essentially indistinguishable from that of traps with fully chaotic dynamics.

We use $^{85}Rb$ atoms in a coherent superposition of their two magnetic-insensitive hyperfine Zeemann ground
states. These two levels, $\left| 5S_{1/2},F=2,m_{F}=0\right\rangle $ denoted $\left| \downarrow \right\rangle $,
and $\left| 5S_{1/2},F=3,m_{F}=0\right\rangle $ denoted $\left| \uparrow \right\rangle $, are separated by the
energy splitting $E_{HF}=\hbar \omega _{HF}$ with $ \omega _{HF}=2\pi \times 3.036$ $s^{-1}$. The atoms are
trapped in a dipole potential formed by a linearly polarized laser close to the $5S_{1/2} \rightarrow 5P_{3/2}$
transition. The dipole potential is inversely proportional to the trap laser detuning $\Delta _{L}$, hence it is
slightly different for atoms in $\left| \uparrow \right\rangle $ and $\left| \downarrow \right\rangle $. The
external (center of mass) potential depends on the internal (spin) state, hence the internal and external degrees
of freedom can not be separated, and the entire Hamiltonian (neglecting interactions between the atoms) is written
as:
\begin{eqnarray}
H&=&H_{\downarrow }\left| \downarrow \right\rangle \left\langle \downarrow \right| +\left( H_{\uparrow
}+E_{HF}\right) \left| \uparrow \right\rangle \left\langle \uparrow \right| \notag \\
&=&\left[ \frac{p^{2}}{2m}+V_{\downarrow }\left( {\bf x}\right) \right] \left| \downarrow \right\rangle
\left\langle \downarrow \right| +\left[ \frac{p^{2}}{2m}+V_{\uparrow }\left( {\bf x} \right) +E_{HF}\right] \left|
\uparrow \right\rangle \left\langle \uparrow \right| \label{haloej}
\end{eqnarray}
where $ V_{\downarrow }$ and $V_{\uparrow }$ are the external potentials for atoms in states $\left| \downarrow
\right\rangle $  and $\left| \uparrow \right\rangle $, respectively. These potentials include the gravitational
potential, equal for both states, and the dipole potential, which can be written as $V_{d,\downarrow }$ and
$V_{d,\uparrow }=(1+\epsilon)V_{d,\downarrow }$, where $\epsilon\equiv\omega_{HF}/\Delta_L$ is the ``perturbation
strength'' typically $10^{-3}-10^{-2}$ in our experiments \cite{Andersen03}.

The eigenenergies of this Hamiltonian consists of two manifolds (belonging to $\left| \downarrow \right\rangle $
and $\left| \uparrow \right\rangle $) separated in energy by $E_{HF}$. The atoms are initially prepared in their
internal ground state $\left| \downarrow \right\rangle $ and their total wavefunction can be written as $\Psi
=\left| \downarrow \right\rangle \otimes $ $\psi $, where $\psi $ represents the center of mass part of their
wavefunction. The echo sequence consists of three short and strong MW pulses, each of which changes the internal
state of the atoms, while leaving the center of mass part of the wavefunction unchanged \cite{Andersen03}. First a
$\pi /2$-pulse puts the atoms into a coherent superposition of $\left| \downarrow \right\rangle $ and $\left|
\uparrow \right\rangle $. After a time $\tau $ a $\pi $-pulse inverts the populations and after another time $\tau
$ the atoms are irradiated by a second $\pi /2$-pulse. The populations of $\left| \downarrow \right\rangle $ and
$\left| \uparrow \right\rangle $ are then measured. If we start with an eigenstate $\left| n_{\downarrow
}\right\rangle $\ of $H_{\downarrow }$ then the population of $\left| \uparrow \right\rangle $ after the echo
pulse sequence is \cite{Andersen03}: $ P_{\uparrow }=\frac{1}{2}\left[ 1- \mathop{\rm Re} \left( F_{echo} \right)
\right] $, where $F_{echo}=\left\langle n_{\downarrow }\left| \eexp{i H_{\downarrow} \tau } \eexp{i
H_{\uparrow}\tau } \eexp{-i H_{\downarrow} \tau } \eexp{-i H_{\uparrow}\tau } \right| n_{\downarrow
}\right\rangle$ is denoted the ``echo amplitude'' (we omit $\hbar$ to simplify the equations). $F_{echo}=1$
indicates perfect coherence and yields $P_{\uparrow }=0$ and $F_{echo}=0$ yielding $ P_{\uparrow }=0.5$ indicates
complete loss of coherence. If $\epsilon=0$ then internal and center of mass motion degrees of freedom decouple
and $F_{echo}=1$ for all times. When considering eigenstates the echo amplitude can be written as a time
correlation function $F_{echo}=\eexp{i \omega_{n,\downarrow}\tau}\left\langle \varphi _{n}\left( t=0\right) \mid
\varphi _{n}\left( t=\tau \right) \right\rangle $ where $\left| \varphi _{n}\left( t=0\right) \right\rangle \equiv
\eexp{-i H_{\uparrow } \tau}\left| n_{\downarrow }\right\rangle $ and $\left| \varphi _{n}\left( t=\tau \right)
\right\rangle \equiv \eexp{-i H_{\downarrow } \tau} \left| \varphi _{n}\left( t=0\right) \right\rangle $.
Therefore, the decay of the echo coherence corresponds to a decay of quantum correlations due to dynamics in the
trap.

In our experiment $^{85}Rb$ atoms are loaded into a far off resonance optical trap, cooled to a temperature of 20
$\mu K$, and optically pumped into the F=2 hyperfine state. By changing the detuning of the trap laser, and
simultaneously adjusting its power, the perturbation strength is controlled. After the MW echo pulses $P_{\uparrow
}$, the population of state $\left| \uparrow \right\rangle $, is measured using fluorescence detection, and the
signal is normalized to $ P_{\uparrow }$ after a short $\pi$-pulse.

The trap is a light-sheet wedge billiard, made from two crossed blue detuned light sheets defining the billiard
walls and where gravity confines the atoms in the vertical direction \cite{Davidson95}. The light sheets have
($1/e^2$) dimensions of $20 \times 250$ $\mu m$, and by the use of cylindrical lenses mounted on rotational
stages, the wedge angle are adjusted in order to control the classical dynamics. The very elongated shape of the
trap allows us to consider only the transverse motion and neglect the longitudinal one, which has a timescale much
longer than the experiment time. The temperature of the atoms is much larger than the mean level spacing in the
trap, and the atoms typically occupy many (up to $\sim10^8$) states in the trap. The measured echo signal is the
ensemble average of all of them.
\begin{figure}[tbp]
\includegraphics[width=3in] {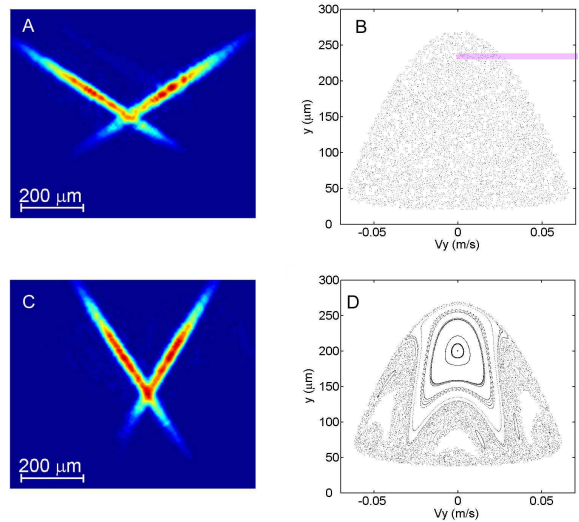}
\caption{(Color online) CCD images of edge billiards with $\alpha=52.5^\circ$ (A) and $\alpha=31^\circ$ (C), used
in our experiments and their corresponding calculated Poincar\'{e} surface of sections, respectively, indicating
chaotic dynamics (B) and mixed dynamics with large islands of stability (D).} \label{p1}
\end{figure}

The structure of phase space in the wedge billiard can be tuned from stability to chaos by varying the vertex
half-angle, $\alpha$ \cite{Wedge}. For $\alpha < 45^\circ$ phase space is mixed, and the size of the stable
islands oscillates as a function of $\alpha$, a behavior dubbed ``breathing chaos''. For $\alpha > 45^\circ$ the
system is fully chaotic. Figures 1B and 1D present the Poincar\'{e} surface of section for the two wedge billiards
of Fig. 1A and 1C, respectively, used in our experiments, calculated by numerical integration of the classical
trajectories in the measured billiard potentials. As seen, the classical dynamics is indeed almost fully chaotic
and mixed for $\alpha = 52.5^\circ$ and $\alpha = 31^\circ$, respectively, as predicted \cite{Wedge} and
previously measured \cite{BilliardsRaizen}.

As shown in Ref. \cite{Andersen04b}, the echo amplitude for small perturbations can be written as:
\begin{eqnarray}
F_{ECH}\left( \left| n_{\downarrow }\right\rangle ,\tau \right)  \simeq 4\sum_{m\neq n}\exp \left( -i\omega
_{m}\tau \right) \left| \left\langle
m_{\uparrow }\mid n_{\downarrow }\right\rangle \right| ^{2}-  \notag \\
\sum_{n\neq m}\exp \left( -2i\omega _{m}\tau \right) \left| \left\langle m_{\uparrow }\mid n_{\downarrow
}\right\rangle \right| ^{2}+\left| \left\langle n_{\uparrow }\mid n_{\downarrow }\right\rangle \right| ^{6},
\label{shex}
\end{eqnarray}
This means that the perturbative echo signal for a monoenergetic ensemble of atoms can be expressed as a function
of the local density of states (LDOS). The LDOS denotes the local average of the absolute value squared of the
matrix elements of the transformation matrix from eigenstates of $H_{\downarrow }$ to eigenstates of $H_{\uparrow
}$. Formally, it is simply $\left| \left\langle m_{\uparrow }|n_{\downarrow }\right\rangle \right| ^{2}$ as a
function of $m_{\uparrow }-n_{\downarrow }$ averaged over a ensemble of neighboring $n_{\downarrow }$ with
approximately the same energy. The width over which the LDOS is nonvanishing is denoted the ``bandwidth'', and if
it is large, we  expect a rapid decay of the echo coherence. In the perturbative regime the LDOS displays system
specific features, despite the fact that the underlying classical dynamics is chaotic. In particular it is evident
from the semiclassical calculations of \cite{Andersen04b,Cohen01} that for atom optics billiards, where the
inherent perturbation is localized on the billiard walls, the LDOS will have pronounced peaks for $ m_{\uparrow
}-n_{\downarrow }$ corresponding to $E_{n}-E_{m}=h/\tau _{bl}$, where $\tau _{bl}$ is the typical time between
encounters with the wall ($\simeq15$ ms in our experiments). This means that the echo amplitude will show partial
revivals for $ \tau =\tau _{bl}$ \cite{Andersen04b}.

\begin{figure}[tbp]
\includegraphics[width=3in] {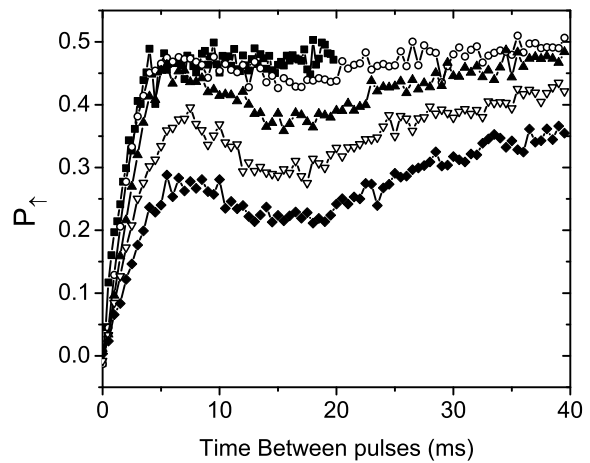}
\caption {Echo signal for a light sheet wedge with chaotic classical dynamics, for different perturbation
strengths (trap laser wavelength tuned from $\lambda=775.9$ to $\lambda=779.7$). $\blacklozenge $: $\epsilon=1.44
\times 10^{-3}$, $\triangledown $: $\epsilon=1.90 \times 10^{-3}$, $\blacktriangle $: $\epsilon=2.43 \times
10^{-3}$, $\circ$: $\epsilon=3.80 \times 10^{-3}$ and $\blacksquare $: $\epsilon=1.52 \times 10^{-2}$. For small
perturbations a nonmonotonic decay with revivals around $\tau=\tau_{bl}$ is seen, whereas for large ones a
monotonic decay is observed.} \label{fi1}
\end{figure}

In Fig. \ref{fi1}, the decay of the echo signal for different perturbations is presented for a wedge with
$\alpha=52.5^\circ$ (see Fig. \ref{p1}A) where the dynamics is almost fully chaotic, as shown in Fig. \ref{p1}B.
For small perturbations a nonmonotonic decay is seen with a partial revival of correlations for $\tau \simeq \tau
_{bl}$ as predicted in \cite{Andersen04b}. The revivals are seen despite the fact that due to the ``high''
temperature  a large band of energies is occupied in the trap. Partial revivals of correlations  in traps where
the dynamics is separable are generally expected at time scales equivalent to the 1D level spacing, but their
surprising observation here, for a trap with chaotic motion, demonstrates that they are a far more widespread
phenomena.

\begin{figure}[tbp]
\includegraphics[width=3in] {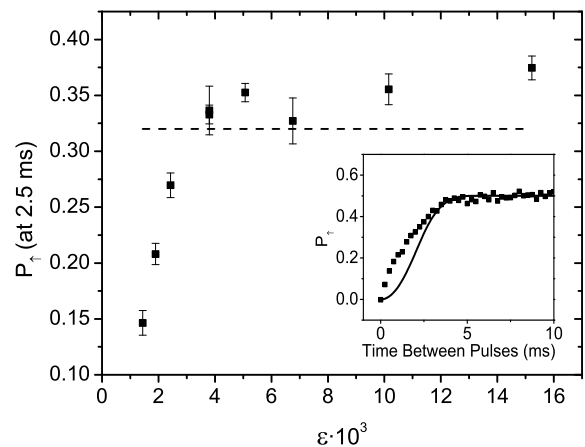}
\caption{Echo signal for a time between pulses of 2.5 ms, as a function of perturbation strength. For a small
perturbation the decay depends strongly on the perturbation strength, whereas for $\epsilon > 4\times 10^{-3}$ the
decay is almost independent of perturbation strength. Dashed line: $P_{\uparrow}$(at 2.5 ms) predicted by the
simple classical model described in the text. Inset: Solid line: $P_{\uparrow}$ calculated by classical model.
$\blacksquare $: $P_{\uparrow}$ measured for a perturbation of $\epsilon=1.52\times 10^{-2}$.} \label{fi2}
\end{figure}
For larger perturbations a crossover to a regime where the decay is monotonic is observed. In this regime the
decay is independent of perturbation strength. This is evident from Fig. \ref{fi2}, in which $ P_{\uparrow }\left(
\tau =0.0025s\right) $ is plotted as a function of perturbation strength. It is seen that $P_{\uparrow }$
initially grows with perturbation strength but for $\epsilon>0.004$, $P_{\uparrow } $ is roughly constant. The
perturbation independent regime of the echo decay is associated with the nonuniversal regime of \cite{Cohen01},
where the LDOS was found to be perturbation independent. In this regime the perturbation is large enough so the
overlap of equivalent eigenstates is small, and this indicates that the effects of quantization of the trap levels
should not play a role and a classical description might be possible. Since the echo amplitude can be viewed as a
propagator, and in our system the thermal de-Broglie wavelength is much smaller than the billiard's dimensions, it
is possible to use a semiclassical propagator \cite{Gutzwiller90}. It is then seen that the classical trajectories
contributing to the ensemble average of the echo amplitude, are those that after evolving forward in time in
$H_{\uparrow}$ and $H_{\downarrow}$, and then backwards in time in $H_{\uparrow}$ and $H_{\downarrow}$, return to
the vicinity of their initial position. These trajectories we divide into two types: those that during this
propagation have hit the wall, and those that did not. Since $H_{\uparrow}$ and $H_{\downarrow}$ are highly
different mainly in the vicinity of the wall, then the action integral along the first type of trajectories will
yield a very large phase and the contribution from these trajectories to the ensemble average of the echo
amplitude will average out. However since the second type of trajectories does not feel the difference between the
potentials it will retrace its forward propagation backwards in time causing the action integral to vanish. These
trajectories will give a perfect contribution. Therefore the echo amplitude in this regime simply measures the
probability that the particles have not yet hit the wall, and this is a classical quantity. The dashed line in
Fig. \ref{fi2} describes the classical estimation of $P_{\uparrow}\left( \tau =0.0025s\right)$ calculated assuming
an idealized hard wall wedge populated with a thermal ensemble of atoms at a temperature of 20 $\mu K$, clipped at
an energy equal to the depth of the trap. In the inset of Fig. \ref{fi2} the classical calculation is shown
together with the measured echo decay for a perturbation of $\epsilon=1.5\times 10^{-2}$, and reasonable agreement
is seen despite the extreme simplicity of the above model.

\begin{figure}[tbp]
\includegraphics[width=3in] {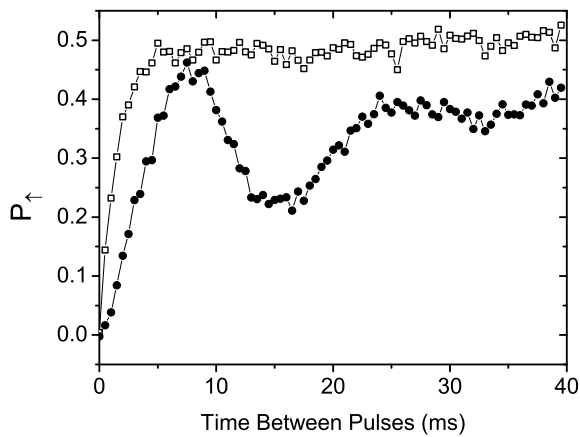}
\caption{Echo signal for a light sheet wedge where the classical phase-space is mixed. $\bullet$: $\epsilon=1.44
\times 10^{-3}$. $\square $: $\epsilon=1.52 \times 10^{-2}$. For the small perturbation a revival that is more
pronounced than the one in the trap with classical chaotic dynamics is observed, whereas for large perturbations
the decay can not be distinguished from the decay in a trap with chaotic dynamics.} \label{fi3}
\end{figure}
Next we consider billiards with mixed dynamics.  The semiclassical perturbative calculations of
\cite{Andersen04b,Cohen01} indicate that classical periodic motion as in elliptic islands will lead to narrow
peaks in the LDOS, thereby yielding more pronounced revivals compared with the chaotic motion discussed above.
This is observed in Fig. \ref{fi3}, where the decay of the echo signal is seen for the mixed-dynamics billiard
with $\alpha=31^\circ$, shown in Fig. \ref{p1}. The revival for small perturbation is deeper than for the chaotic
motion of Fig. \ref{fi1}, and the reminiscence of a second revival for $\tau \simeq 2\tau _{bl}$ can be seen
indicating that the peak in the LDOS is narrower. Since the revivals are more pronounced in the case of mixed
phase space they ``survive'' to slightly larger perturbation strength. However, in the perturbation-independent
regime also shown in Fig. \ref{fi3}, the decay is essentially indistinguishable from that of the chaotic billiard,
in agreement with the classical model given above.

In summary, we studied the decay of quantum correlations in atom optics billiards in which the classical dynamics
is chaotic or mixed. We observed two distinct regimes for the perturbation strength in which the decay was
qualitatively different. In the perturbative regime the decay was non-monotonic, with revivals at times
corresponding to the typical time between bounces from the wall. The revivals were more pronounced in traps with
mixed phase-space as compared with traps where the dynamics is almost fully chaotic. In the perturbative regime
the decay rate increased with perturbation strength. However, in the perturbation independent regime the decay was
monotonic and independent of perturbation strength. No clear distinction between the decay in traps with mixed and
chaotic dynamics could be made in the perturbation independent regime. In this regime the echo amplitude measures
the classical probability that an atom have not yet hit the wall.

The authors gratefully acknowledge discussions with U. Smilansky, D. Cohen, T. Kottos and K. Molmer. This work was
supported in part by the Israel Science Foundation and the Minerva Foundation.


\begin{references}

\bibitem{Zurek03} W.H. Zurek., Rev. Mod. Phys. {\bf 75}, 715 (2003).

\bibitem{DiVincenzo95} D.~P. DiVincenzo, Science {\bf 270}, 255 (1995).

\bibitem{Gutzwiller90} M.~C. Gutzwiller. \newblock {\it Chaos in Classical and Quantum Mechanics}, Springer-Verlag, New York (1990).

\bibitem{Peres84}  A. Peres, Phys. Rev. A {\bf 30}, 1610 (1984).

\bibitem{Jalabert01}  R. A. Jalabert and H. M. Pastawski, Phys. Rev. Lett.,
{\bf 86}, 2490 (2001).

\bibitem{Emerson02} J. Emerson, Y. S. Weinstein, S. Lloyd, and D.
G. Cory, Phys. Rev. Lett. {\bf 89, } 284102 (2002); Ph. Jacquod, I. Adagideli, and C. W. J. Beenakker, Phys. Rev.
Lett. {\bf 89}, 154103 (2002); F.M. Cucchietti, H.M. Pastawski, and D. A. Wisniacki, Phys. Rev. E {\bf 65},
045206(R) (2002); T. Prosen and M. Znidaric, New J. Phys. {\bf 5} 109 (2003); R. Sankaranarayanan, and A.
Lakshminarayan, Phys. Rev E {\bf 68} 036216 (2003), M.~Hiller, T.~Kottos, D.~Cohen, and T.~Geisel., Phys. Rev.
Lett. {\bf 92}, 010402 (2004).

\bibitem{Raizen99} M.~Raizen, Adv. At. Mol. Opt. Phys. {\bf 41}, 43 (1999).

\bibitem{Tunneling} D.~A. Steck, W.~H. Oskay, and M.~G. Raizen, Science {\bf 293}, 274 (2001); W.~K. Hensinger et. al., Nature {\bf 412}, 52 (2001).

\bibitem{Accelerator} M.~Oberthaler et. al., Phys. Rev. Lett. {\bf 83}, 4447 (1999); S.~Schlunk, M.B. ~d'Arcy, S. A. ~Gardiner, and G. S. ~Summy, Phys. Rev. Lett. {\bf 90}, 124102 (2003).

\bibitem{BilliardsRaizen} V. Milner, J. L. Hanssen, W. C. Campbell and M. G. Raizen, Phys. Rev. Lett. {\bf 86}, 1514
(2001).

\bibitem{Billiards} N. Friedman, A. Kaplan, D. Carasso and N. Davidson, Phys. Rev. Lett. {\bf 86}, 1518 (2001); A. Kaplan, N.
Friedman, M. Andersen, and N. Davidson, Phys. Rev. Lett. {\bf 87}, 274101 (2001).

\bibitem{Andersen03}  M. F. Andersen, A. Kaplan and N. Davidson, Phys. Rev. Lett. {\bf 90}, 023001 (2003).

\bibitem{Andersen04b} M.~F. Andersen, T.~Grunzweig, A.~Kaplan, and N.~Davidson, Phys. Rev. A, {\it in press} (2004).

\bibitem{Davidson95}  N. Davidson et. al., Phys. Rev. Lett. {\bf 74,} 1311 (1995).

\bibitem{Wedge}  H. E. Lehtihet and B. N. Miller, Physica (Amsterdam) {\bf 21D}, 93 (1986). 

\bibitem{Cohen01}  D. Cohen and T. Kottos, Phys. Rev. E, {\bf 63}, 036203 (2001); D. Cohen, A. Barnett and E. J. Heller, Phys. Rev. E, {\bf 63}, 046207 (2001).



\end{references}
\end{document}